\documentclass[twocolumn,aps,pre,superscriptaddress]{revtex4}
\usepackage{hyperref}
\usepackage{amsmath, amssymb, graphicx, float, xcolor}
\usepackage{bm}

\graphicspath{{../../Figures/}}

\begin{document}

\title{Resource competition and survival for random walkers on complex networks}


\author{Snehal M. Shekatkar}
\email[]{snehal@inferred.in}
\affiliation{School of Computing and Data Sciences, FLAME University, Pune 412115, India}



\begin{abstract}
We consider the problem of $W$ random walkers on a complex network where each walker requires an amount $\theta$ of resource at each time to survive. Resource is stochastically produced at each vertex of the network with a specified distribution. The survival time of the system of walkers is defined as the time required for all $W$ walkers to die. Using numerical simulations, we show that Erd{\H o}s-R\'enyi networks a have higher mean survival time than Scale-free networks as expected, but also that uniform distribution production leads to higher mean survival time than exponential distribution with the same mean. We derive an exact expression for the survival-time distribution under the assumption of independent walker lifespans, and use it to show that scale-free network topology produces strong dependence in the lifespans but homogeneous topology does not. We moreover demonstrate how the survival time can be substantially increased without changing the resource production or network topology by appropriately biasing the movement of the walkers towards low degree vertices on Scale-free networks, but has no effect on Erd{\H o}s-R{\'e}nyi networks.
\end{abstract}

\maketitle



Random walks on complex networks provide a simple paradigm to study various transport phenomena taking place on networks \cite{Lovasz1996, NohRieger2004, MasudaPorterLambiotte2017}. It has been well-studied, and classic results include the stationary distribution, computation of mean first-passage times \cite{Redner2001, CondaminEtAl2007}, probability of extreme events \cite{KishoreEtAl2012, GandhiSanthanam2022}, etc. Many themes of the models have been studied as well which include biased random walks, walks on weighted networks \cite{ZhangShanChen2013, RiascosMateos2021}, and even walks on temporal networks \cite{StarniniEtAl2012}. Although the main goal of the model is to understand the effect of the network structure on the transport phenomena taking place on it, it has also been used to probe the network structure itself. For example, random walks have been used to detect community structure in networks \cite{RosvallBergstrom2008}, and also to define a new type of centrality for vertices in simple and multilayer networks \cite{NohRieger2004, SoleRibaltaEtAl2016}. Apart from the main academic interest, random walks on networks have found a large number of applications in the real-world. For example, they can be used to study diffusion and percolation in disordered systems \cite{BenAvrahamHavlin2000}, navigability in transportation networks \cite{DeDomenicoEtAl2014, GallottiBarthelemy2014}, and for network search \cite{PageBrinMotwaniWinograd1999}.

It is natural to use random walks to model movement of various animals from one region to other in search of resources like food and water. Their survival critically depends on not too many individuals competing for the resources at the same location. The same idea could be applied to human societies whereby too much crowding of a region leads to each person getting smaller share of available resources on average. To turn this into a mathematical model, we consider random walks on networks where walkers consume resources on the vertices of an undirected network. We find that Erd{\H o}s-R\'enyi networks have substantially higher survival time than the Scale-free networks with the same average degree. As we later show, on Scale-free networks, the survival time can be significantly improved by biasing the movement of the walkers towards low degree vertices. We also show that this strategy has no effect on Erd{\H o}s-R\'enyi networks. 

\section{\label{sec:model} The model}
We consider a discrete-time random walk on a network with $W(0)$ number of initial walkers distributed uniformly randomly on its $N$ vertices. Each vertex $v$ of the network stochastically generates an amount $X_v(t)$ of a \emph{perishable} resource at each discrete time $t$, available for consumption by the walkers. Each walker needs an amount $\theta$ to survive at each time step which it can procure only from its current vertex. Perishability implies that walkers cannot consume leftover amounts from the past, and must rely on the current amounts. At each time step $t$ three things happen:
\begin{enumerate}
    \item{Each of the $W_u(t)$ walkers on vertex $u$ moves to a randomly chosen adjacent vertex $v$ with probability proportional to $k_v^{\eta}$ where $k_v$ is the degree of $v$ and $\eta$ is the bias parameter.}
    \item{Vertices then produce resource amounts $X_u(t)$ for $u=1, 2, \dots, N$ which are i.i.d. random variables with a given distribution $p(x)$, henceforth called the \emph{resource kernel}.} 
    \item Each of the $N_v(t)$ walkers on vertex $v$ at time $t$ receives the amount $Y_v(t) = X_v(t)/N_v(t)$, and if $Y_v(t) < \theta$, all the walkers on $v$ die. 
\end{enumerate}
We define the survival time $T_s$ of the system of walkers as the time at which the last walker dies:
\begin{equation}
    \begin{aligned}
    T_s = \min\{t\in \mathbb{N}\ |\ W(t) = 0\}
    \end{aligned}
    \label{eq:}
\end{equation}
Notice that this means that as long as $W(0) > 0$, the smallest value of $T_s$ is $1$.
We are interested in understanding how the mean survival time $\langle T_s\rangle$ depends on $W(0)$, the network topology, the resource kernel $p(x)$, the threshold $\theta$, and the bias $\eta$. 

\subsection{Independent walker lifespans approximation}
Clearly, the lifespans of individuals walkers are not independent since two or more walkers dying on the same vertex have the exact same lifespan. However, assuming that all lifespans are independent allows us to find the exact distribution of survival time. This would be approximately true when $W(0) << N$, and if the walkers are uniformly randomly distributed on the network since the chance of two or more walkers dying on the same vertex is negligible then.
In this case, the probability that the lifespan $\tau_i$ of walker $i$ is exactly $t$ is: 

\begin{equation}
    \begin{aligned}
    \mathbb{P}(\tau_i = t) = (1-F(\theta))^{t-1}F(\theta)
    \end{aligned}
    \label{eq:}
\end{equation}
where $F$ is the CDF of the kernel $p(x)$. This equation says that for a walker to die exactly at time $t$, it must survive at all previous $t-1$ time steps, and then it dies at time $t$ with probability $F(\theta)$. 
Hence, 
{\small
\begin{equation*}
    \begin{aligned}
    \mathbb{P}(\tau_i \leq t) = F(\theta)\sum\limits_{t^{\prime}=1}^t (1-F(\theta))^{t^{\prime}-1} = F(\theta)\sum\limits_{t^{\prime}=0}^{t-1} (1-F(\theta))^{t^{\prime}}
    \end{aligned}
    \label{eq:}
\end{equation*}
}
which gives:
\begin{equation}
    \begin{aligned}
    \mathbb{P}(\tau_i \leq t) = 1-(1-F(\theta))^{t}
    \end{aligned}
    \label{eq:}
\end{equation}
Notice that: 
$$T_s = \max\{\tau_1, \tau_2, \dots, \tau_{W(0)}\}$$ 
To compute the distribution of $T_s$, let $\phi(t)$ be its CDF at $t$. Then we have:
\begin{equation*}
    \begin{aligned}
    \phi(t) &= \mathbb{P}(T_s \leq t) = \mathbb{P}(\max\{\tau_1, \tau_2, \dots, \tau_{W(0)}\} \leq t) \\
            &= \mathbb{P}(\tau_1 \leq t, \tau_2 \leq t, \dots, \tau_{W(0)} \leq t) 
    \end{aligned}
    \label{eq:}
\end{equation*}
Using independence of $\tau_i$, we can write this as:
\begin{equation}
    \begin{aligned}
\phi(t) &= \prod\limits_{i=1}^{W(0)}\mathbb{P}(\tau_i \leq t)
    = \prod\limits_{i=1}^{W(0)}[1-(1-F(\theta))^{t}] \\
    &= [1-(1-F(\theta))^{t}]^{W(0)}
    \end{aligned}
    \label{eq:}
\end{equation}
which finally gives:
{\small
\begin{equation}
    \begin{aligned}
    \mathbb{P}(T_s = t) &= [1-(1-F(\theta))^{t}]^{W(0)}\\
    &\quad -[1-(1-F(\theta))^{t-1}]^{W(0)}
    \end{aligned}
    \label{eq:independent_walkers_Ts}
\end{equation}
}
This can be used to compute the mean survival time in the independent walkers limit, and provides the upper bound since in general many walkers will meet and die on the same vertex. Also notice that this result is completely independent of the network topology.
\section{Results}
We consider two types of network topologies: Scale-free and Erd{\H o}s-R\'enyi. Both network topologies have the same average degree. The networks are generated using the configuration model as follows. Scale-free networks have degree distribution: 
\begin{equation}
    \begin{aligned}
    \begin{cases}
    p(k) &= k^{-\alpha}/\zeta(\alpha, k_\text{min})\quad \text{for}\ k \geq k_{\text{min}}\\
        &\\
        &= 0\quad \text{Otherwise}
    \end{cases}
    \end{aligned}
    \label{eq:}
\end{equation}
where $\zeta(s, a)$ represents the Hurwitz zeta function: 
\begin{equation}
    \begin{aligned}
    \zeta(s, a) = \sum\limits_{n=0}^{\infty} \frac{1}{(n+a)^s}
    \end{aligned}
    \label{eq:}
\end{equation}
To generate scale-free networks with given values of $\alpha$ and $k_{\text{min}}$, we draw the degrees from this distribution and use stub-matching to generate a configuration model network. To generate an equivalent Erd{\H o}s-R{\'e}nyi network, observe that the average degree of the scale-free configuration model network is given by:
\begin{equation}
    \begin{aligned}
    \langle k\rangle = \sum\limits_{k=k_\text{min}}^{\infty} k^{1-\alpha} / \zeta(\alpha, k_\text{min})
    \end{aligned}
    \label{eq:}
\end{equation}
We then generate an Erd{\H o}s-R\'enyi network by drawing degrees from the Poisson distribution with mean $\langle k\rangle$. It is possible that a network so generated is fragmented into several components, but the choice of our parameter values gives rise to networks with most vertices belonging to a single large component. Throughout the paper, we use $\alpha=2.2$ and $k_{\text{min}} = 2$, which gives us $\langle k\rangle \approx 9.36$. 

We also use two different resource kernels: Uniform and Exponential: $X \sim \text{Exponential}(\lambda)$ and $X \sim U(0, A)$ with $\lambda=1$ and $A=2$, so that in both the case $\langle X\rangle = 1$. We then run the model on these networks with $10000$ initial walkers, and average the results over $1000$ random realizations of network topology. 

\subsection{Variation with threshold for unbiased walks}
\begin{figure}[ht]
    \includegraphics[width=0.8\columnwidth]{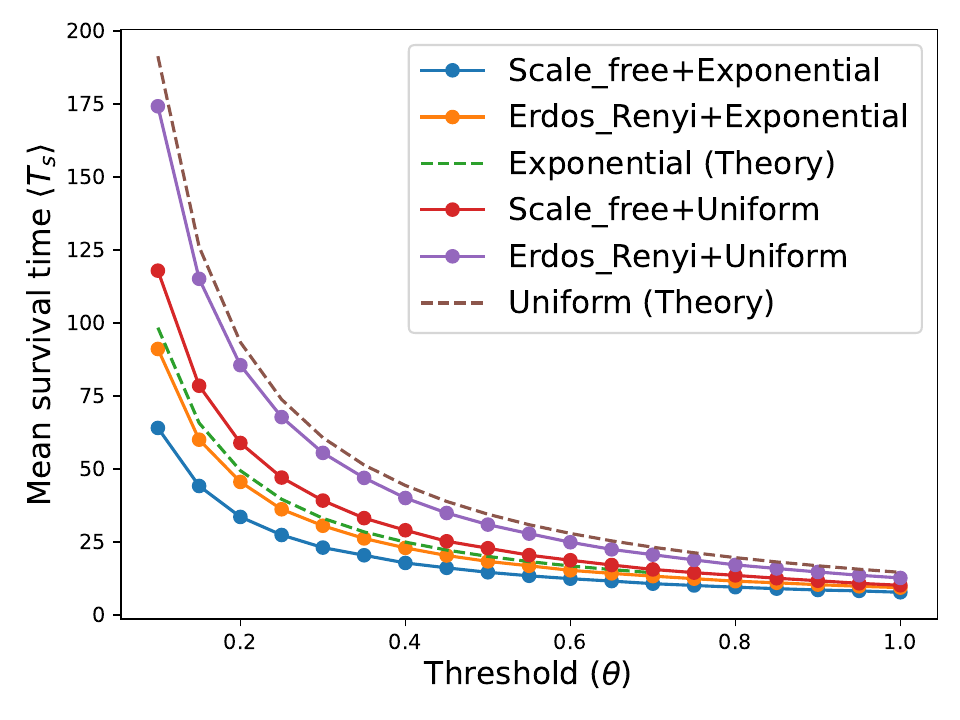}
    \caption{\label{fig:time_vs_threshold} Dependence of the mean survival time on the walker threshold $\theta$. It can be seen that homogenous network structure (ER) and uniform resource production are independently better at increasing the system's survivability. Also, understandably, if each walker requires higher resource amount to survive, walkers tend to die faster reducing $\langle T_s\rangle$. The dotted curves are obtained using Eq(\ref{eq:independent_walkers_Ts}). All results are averaged over $1000$ random realizations on networks of size $N=10000$ each. Other parameters are stated in the text.}
\end{figure}

We first want to see how the mean survival time depends on the threshold $\theta$ when the walk is unbiased ($\eta=0$). Fig.~\ref{fig:time_vs_threshold} shows the variation of survival time on the threshold for all four combinations of network topology and resource generation distribution. It is no surprise that $\langle T_s\rangle$ decreases with increase in $\theta$ as more amount is required for survival, but two things are worth noticing. First, Erd{\H o}s-R{\'e}nyi network performs better than Scale-free in terms of keeping the walkers alive longer although both have the same mean degree. The reason is that the chances of survival go down on high-degree vertices, and there are too many of those in a Scale-free network. Thus, hubs in Scale-free networks kill the walkers very fast. The same is true for uniform distribution compared to the exponential distribution although both have the same mean $1$. It is also reasonable that at high values of $\theta$, both network topology and the form of resource generation do not matter, and all four combinations converge to the same value. The dotted curves in the same figure are obtained by using Eq(\ref{eq:independent_walkers_Ts}). From this, Erd{\H o}s-R\'enyi network can be seen to be treating the walkers to be almost independent, but Scale-free network creates a strong dependence in their movement and significantly brings down the survival time.
\begin{figure}
    \includegraphics[width=0.9\columnwidth]{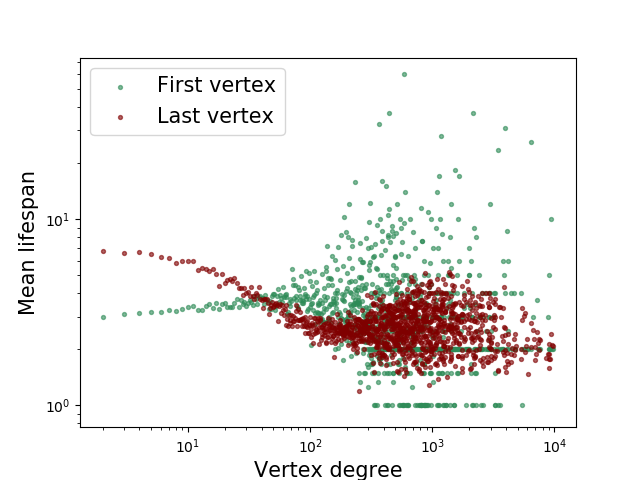}
    \caption{\label{fig:lifespan_vs_deg} Mean lifespan of a walker as a function of the degree of its first and last vertex on the Scale-free network with uniform resource kernel ($\langle X\rangle = 1$) and $\theta=0.1$. As the plot shows, it is better for a walker to start its journey on a high-degree vertex. Also, if a walker dies on a high-degree vertex, it implies that the walker did not survive long. The plot is noisy for high-degree vertices because there are relatively few of those compared to low-degree ones. Results are obtained by taking $100$ random realizations of networks of size $N=10000$.}
\end{figure}

To gain deeper understanding of the process, we can look at the dependence of average lifespan of a walker on its first and last vertex. Thus, we look at the average lifespan of a walker who starts at vertex with degree $k$, as a function of $k$. The green dots in Fig.~\ref{fig:lifespan_vs_deg} show this variation for Scale-free networks. Similarly, we look at the average lifespan of a walker which dies on a vertex with degree $k$, as a function of $k$. The red dots in the same figure show this variation. Thus, we see that it is better for a walker to start on a high-degree vertex. Similarly, if we know that a walker died on a high-degree vertex, then it is more likely that they did not survive for long. 

\begin{figure}
    \includegraphics[width=0.8\columnwidth]{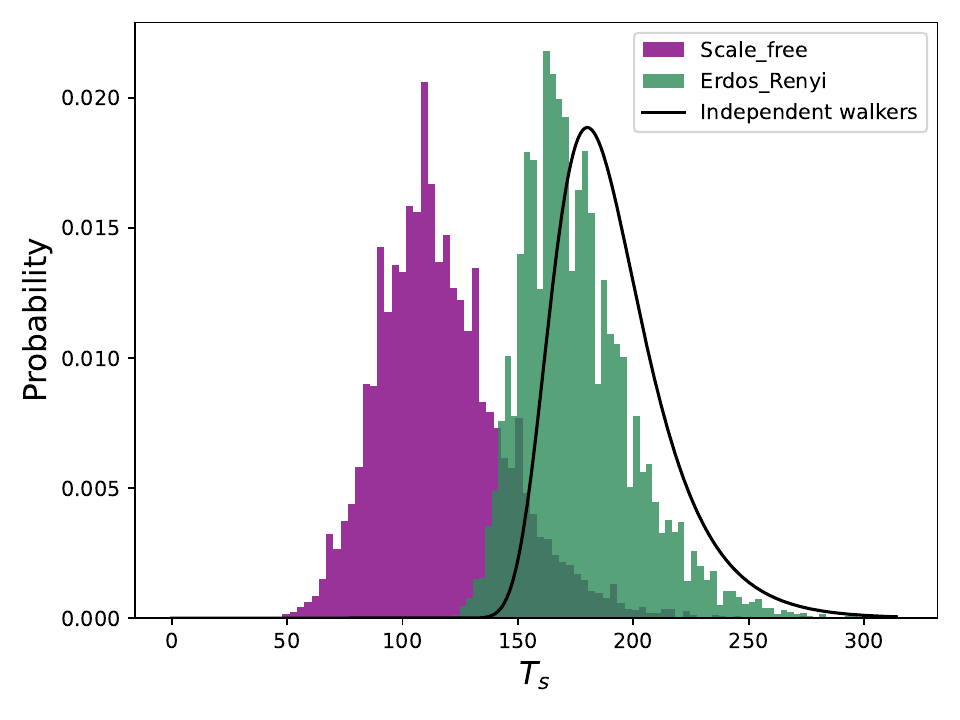}
    \caption{\label{fig:survival_times_hist} Distributions of the survival time for Erd{\H o}s-R\'enyi and Scale-free networks for $\eta = 0.0$ and $\theta=0.1$. The resource kernel is uniform with $\langle X\rangle = 1$. The black curve shows the distribution obtained under the assumption of walkers' movements being independent (Eq(\ref{eq:independent_walkers_Ts})). The plot conveys that the Scale-free topology generates strong interdependence in the movement of walkers, but homogenous topology of Erd{\H o}s-R\'enyi doesn't.}
\end{figure}
\subsection{Improving the survival time by degree-biasing}
\begin{figure}[ht]
    \includegraphics[width=0.8\columnwidth]{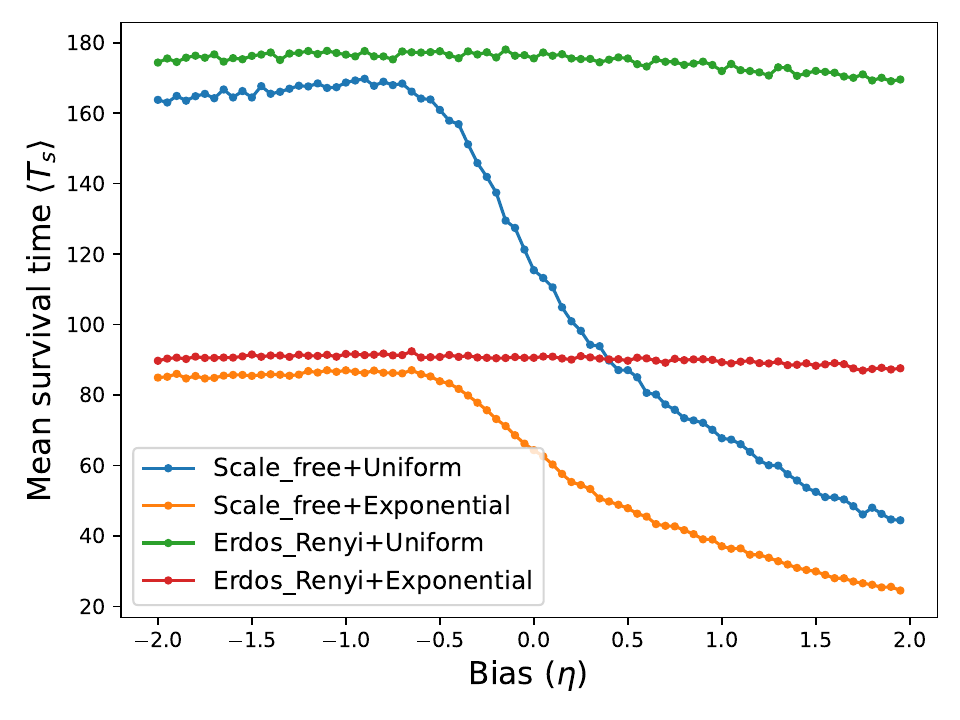}
    \caption{\label{fig:ave_time_vs_bias}Mean survival time of the system as a function of the movement bias for all four combinations of network topology and resource production with $\theta=0.1$ and $\langle X\rangle = 1$. It can be seen that bias towards low-degree vertices ($\eta < 0$) substantially increases the survival time on Scale-free topology, but any bias (positive or negative) has no effect for the Erd{\H o}s-R\'enyi topology. Parameters: $N=10000$, $W(0)=10000$. Results are averaged over $1000$ random realizations.} 
\end{figure}
An interesting question is whether the mean survival time for a fixed value of $\theta$ can be increased without changing the network structure or increasing the mean resource production on the vertices. We here show that this is possible for scale-free networks by biasing the random walkers towards low-degree neighbouring vertices. We can also explore the consequences of biasing them towards high-degree neighbours. To do this, notice that the probability of choosing a particular neighbour $v$ is proportional to $k_v^{\eta}$ where $k_v$ is neighbour's degree. For negative $\eta$, the probability is larger for low-degree vertices. In Fig.~\ref{fig:ave_time_vs_bias}, we show the variation of $\langle T_s\rangle$ with $\eta$. As the plot shows, in Scale-free networks, the survival time significantly increases, and reaches close to the value of corresponding ER case. However, Erd{\H o}s-R\'enyi network is completely indifferent to this strategy. This is understandable since ER has homogeneous structure. 

\section{Conclusion}
We have proposed a simple random walk model of the effect of overcrowding of individuals on their survival when limited amount of resource is available. We showed that three things independently affect the survivability of walkers: the network topology, probability distribution of generation of resources, and the bias of walkers towards the neighbouring vertices. In particular, Erd{\H o}s-R\'enyi network can support walkers for far longer than networks with heavy-tailed distributions. Similarly, the model shows that uniform random resource production outperforms the exponential random production even when they have the same means. Finally, we demonstrated how the survivability of the walkers can be significantly improved in scale-free networks simply by tweaking the movement of the walkers without changing the network structure or the resource production. It also shows that on homogeneous networks like the Erd{\H o}s-R\'enyi network, the biased random walk is as good as unbiased one in terms of survivability, and the only way to improve is to increase the mean resource production. Our work sheds light on mechanisms that can help us in improving resource sharing in the real-world systems, and opens up several interesting questions such as what happens to survivability when walkers are non-identical, when resource productions are correlated in connected vertices, and so on. We plan to focus on some of these questions in future work.


\end{document}